\documentclass{cernrep} 
\usepackage{texnames} \usepackage{xfrac}
\usepackage[T1]{fontenc}
\usepackage[bookmarks, colorlinks=true, linktoc=page, pdftex, linkcolor=black, citecolor=black, urlcolor=blue]{hyperref}
\newcommand{\RM}[1]{\mathrm{#1}}
 
\renewcommand{\d}[2]{\frac{d #1}{d #2}} 
 
\usepackage{fancyhdr}
\fancyhfoffset{4 mm}
\fancypagestyle{ARTTITLE}{%
\fancyhf{} 
\lhead{\small{Proceedings of the CERN--Accelerator--School course: \it{Introduction to Accelerator Physics}}} 
\lfoot{Available online at \href{https://cas.web.cern.ch/previous-schools}{https://cas.web.cern.ch/previous-schools}}
\rfoot{\thepage\hspace*{3mm}}
 
}

\begin{document}
\title{Accelerator Vacuum}
 
\author {M.\,Seidel}

\institute{Paul Scherrer Institut and \'Ecole Polytechnique F\'ed\'ederale Lausanne, Switzerland}

\begin{abstract}
This lecture introduces major physics and technology aspects of accelerator vacuum systems. Following an introduction, in the second section generic vacuum quantities such as pressure, gas density, the gas equation, pumping speed, conductance are introduced. Since accelerators typically have lengthy vacuum tubes, one-dimensional calculation is in many cases sufficient to compute a pressure profile for an accelerator, and  methods for doing so are developed in the next section. In the fourth section accelerator specific aspects of vacuum are considered. This includes lifetime limiting effects for the particle beam, such as bremsstrahlung, elastic and inelastic scattering. Requirements for vacuum properties are derived. In the fifth section types of components and suitable materials for accelerator vacuum systems are described. Such components are for example flange systems, vacuum chambers for accelerators and the different types of pumps.
\end{abstract}

 \maketitle 

\thispagestyle{ARTTITLE}
 
\section{Introduction}
\begin{figure} 
\centering\includegraphics[width=0.85\textwidth]{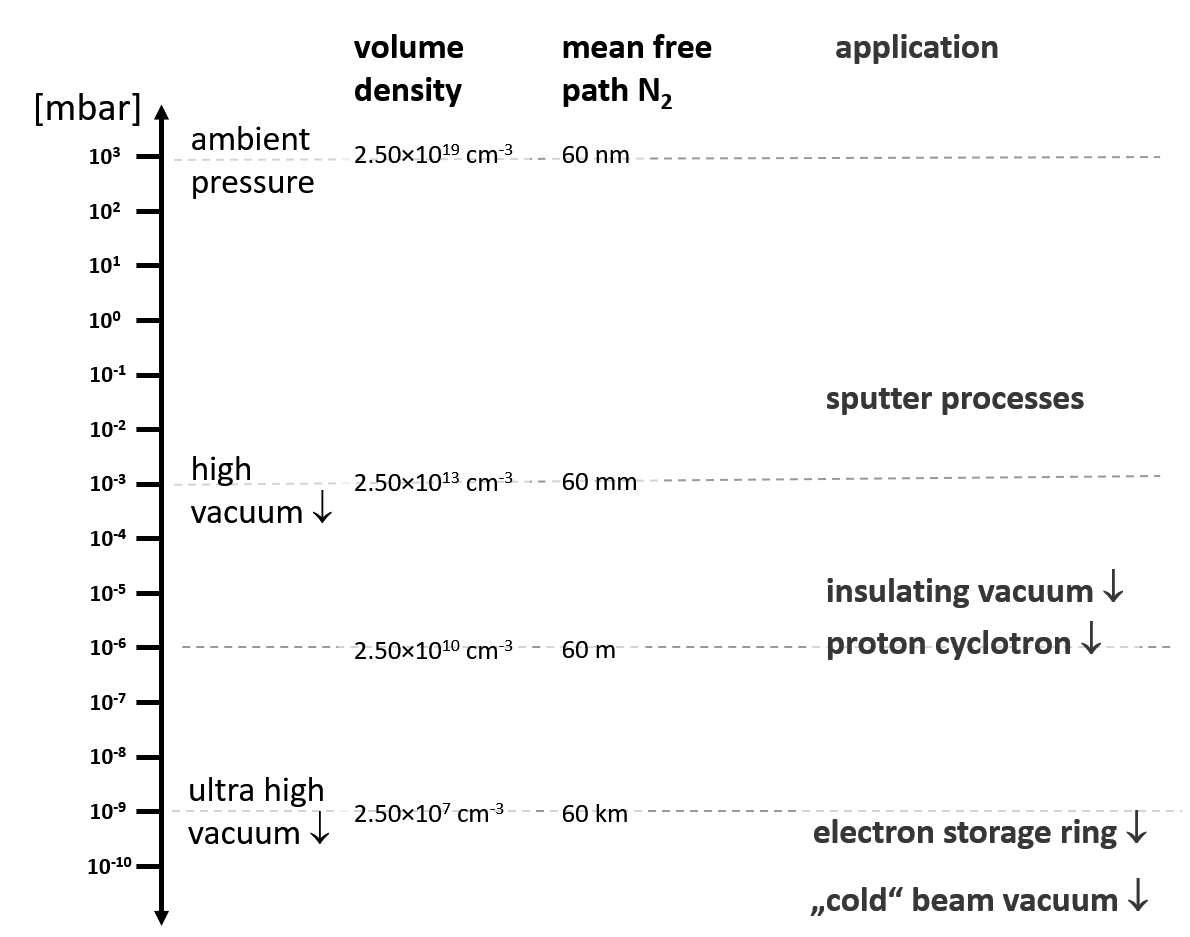}
\caption{Gas pressures ranging from ambient conditions down to cold accelerator vacuum cover 13 orders of magnitude.}
\label{fig:p_overview}
\end{figure}

Every particle accelerator needs a vacuum system to transport its particle beams with low losses. With a~certain probability beam particles interact with residual gas molecules. The scattering leads to immediate particle losses from the beam, or to a deterioration of the beam quality. Both effects are not wanted and the beam losses can even lead to a problematic activation of accelerator components. For different situations, i.e. beam particle species and beam energies, the requirements for the quality of acceptable vacuum conditions can differ significantly. For example in a hadron accelerator with a section that is passed by the beam only once, a pressure of $10^{-5}\,$mbar might be sufficient. On the other hand an~electron storage ring might need a base pressure of $10^{-10}\,$mbar in the absence of beam. Consequently the~technology and physics of vacuum systems covers a broad range of effects and concepts, depending on the application. Vacuum systems for electron and proton storage rings are discussed for example in Ref. \cite{benvenuti}, and a dedicated CERN school on accelerator vacuum is documented under Ref. \cite{CERN}. Fundamentals of vacuum physics are discussed in Ref. \cite{redhead}. 
Figure\,\ref{fig:p_overview} shows the range of pressure levels from ambient conditions down to the lowest pressures in accelerators. The pressure ranges for typical applications, the volume density of molecules and the mean free path of nitrogen molecules are indicated. The pressure range for beam vacuum is in technical language referred to as Ultra-High-Vacuum (UHV). One often speaks about dynamic vacuum, when the presence of an intense beam affects the residual gas pressure.

\section{Pressure and Gas Equation}
Pressure is basically a force that gas molecules apply to a surface by mechanical momentum transfer, averaged over a huge number of collisions during a macrosopic time interval. Pressure is measured as a~force per area. A common unit is Pascal, 1\,Pa~=~1\,N/m$^2$. Another common unit is 1\,mbar~=~$100$\,Pa. The~average velocity of molecules depends on the square root of the temperature as shown in Eq. (\ref{eq:velo}). Using the molecule density $n_v$ the average velocity can be related to the rate of molecules impinging on the wall of a vessel per area and per time.

\begin{equation}
\overline{v} = \sqrt{ \frac{8 k_b}{\pi m_0} ~ T },~~ \d{N}{A\,dt} = \frac{1}{4} n_v \overline{v}
\label{eq:velo}
\end{equation}

Here $k_b = 1.38\times 10^{-23}\,$J/K is the Boltzmann constant. The gas equation is a fundamental law relating pressure, volume, temperature and amount of a gas:

\begin{equation}
PV = N k_b T = nRT.
\label{eq:pv}
\end{equation}

$R = 8.314\,$N\,m\,/\,mole\,K is the gas constant. The majority of vacuum systems operate at room temperature and we can consider the temperature as being constant. Consequently the product of pressure and volume is a measure of the amount of gas, related to the number of molecules $N$ or the number of moles $n$. In practice a leak rate is often given in terms of mbar\,l/s, which is the amount of gas that is entering the considered recipient per unit of time.
Vacuum systems are equipped with pumps, devices that absorb gas molecules. A vacuum pump is characterised by its pumping speed $S = Q/P$, quantified in l/s at its interface. Here $Q$ is the gas load at the interface of the pump. The pumping speed varies for different gas species. Some chemically inactive gases like the inert gases He, Ar, and for example methane (CH$_4$), are pumped less efficiently by several types of pumps like Titanium sublimation pumps and NEG pumps. Turbo pumps and cryo pumps are not based on chemical binding reactions and are suited for all gas types. 
In vacuum systems we can discriminate two types of flow regimes - viscous flow and molecular flow. Viscous flow occurs for higher gas densities. The higher the density of a gas the~shorter is the mean free path between two collisions of a gas molecule with others. The Knudsen number is the ratio between the mean free path and a typical dimension of the vacuum recipient. If the~mean free path is much larger than a typical dimension, we have molecular flow. With the cross section $\sigma$ for collissions the mean free path $\lambda$ is calculated as follows:

\begin{equation}
\lambda = \frac{k_b T}{\sqrt{2} \sigma P}.
\end{equation}

For example Nitrogen residual gas at room temperature and a pressure of 10$^{-6}$\,mbar has a mean free path of 60\,m, which is much larger than the diameter of a typical vacuum pipe. Thus for particle accelerators we normally deal with molecular flow from a source of residual gas to a pump. We define the conductance of a vacuum component, e.g. an orifice or a piece of vacuum tube, as the ratio of the~molecular flow and the pressure drop across the element: 

\begin{equation}
C = \frac{Q}{\Delta P} . 
\label{eq:cond}
\end{equation}

The conductance of an orifice of cross section $A$ can be estimated by:
\begin{equation}
C = \sqrt{\frac{k_b T}{2\pi M}} A\, , ~~~
C_\mathrm{air} = 11.6 \mathrm{[l/s]} ~
A \mathrm{[cm^2]}.
\end{equation}

Orifices may be used to realise a defined pumping speed for outgassing measurements. The conductance of a circular tube with diameter $d$ and length $l$ is given by:
\begin{equation}
C = \sqrt{\frac{2 \pi k_b T}{M}} \frac{d^3}{l} \, , ~
C_\mathrm{air} = 12.1 \mathrm{[l/s]} \,
\frac{d^3 \mathrm{[cm]}}{l \mathrm{[cm]}}.
\end{equation}

Conductance varies with the type of gas, and in both formulas $M$ denotes the molecular mass of the~considered gas species. If two components are connected, the resulting conductance can be calculated from the individual conductances $C_1, C_2$. For a concatenation in series the resulting conductance is obtained by inverse addition: 

\begin{equation}
C_\mathrm{total} = \left( \frac{1}{C_1} + \frac{1}{C_2} \right) ^{-1}.
\end{equation}

And for a combination in parallel, simple addition has to be applied: 

\begin{equation}
C_\mathrm{total} = C_1 + C_2. 
\end{equation}

In a practical example an ion sputter pump of $400\,$l/s is connected to a recipient by a 30\,cm long, $d=8\,$cm tube. This results in an effective reduction of the pumping speed to $136\,$l/s. 

\begin{figure} 
\centering\includegraphics[width=0.90\textwidth]{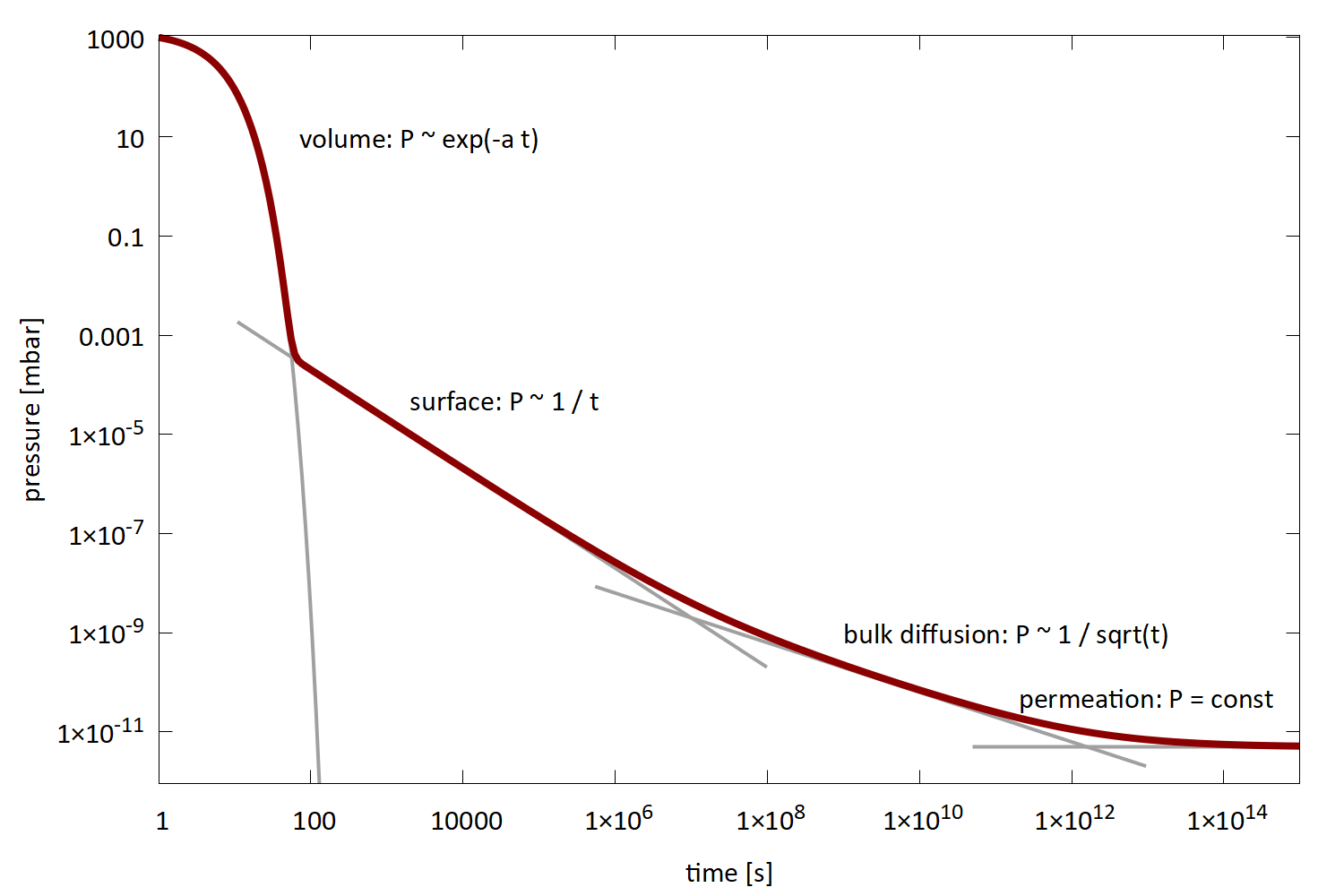}
\caption{Qualitative pump down behaviour of a recipient. Note the large logarithmic time scale on which $10^8$\,s corresponds to three years.}
\label{fig:pdown}
\end{figure}

In an accelerator vacuum system the operating pressure is established as a balance between release rate of free gas molecules and the removal rate that results from conductance and pumping. Gas molecules are released by different effects, such as thermal desorption from surfaces, beam induced desorption, diffusion of gas molecules from the bulk of material, permeation from ambient conditions through material, but also from leak rates through sealed connections between components. Among those effects thermal desorption is typically a significant contribution. Because of thermal energy, physical and chemical bindings of gas molecules to the material of the wall can be broken up and molecules are released. The sojourn time of a molecule at the wall is an exponential function of the temperature: $\tau \propto \exp{\left( E_d\,/\,k_b T \right) }$. For example a binding energy of 1\,eV at room temperature leads to a sojourn time of 5 hours. The qualitative behaviour of the pressure in a recipient during pump-down, that results from the mentioned outgassing mechanisms, is shown in Fig.\,\ref{fig:pdown}. In the first phase of the pumpdown the~amount of gas removed from the recipient per unit time is proportional to the pressure, and thus one observes an exponential pressure decay. At some point a balance is reached between pumping speed and release of gas molecules from the surface of the walls into the volume of the recipient. In this phase the pressure decays inversely with time until the surface is emptied. Afterwards the outgassing is dominated by molecules that are transported in the bulk material to the surface by thermal diffusion. According to the~nature of diffusion this process scales as $1/\sqrt{t}$. It is possible to reduce this contribution to outgassing by a heat treatment of the material under vacuum conditions, thereby enhancing diffusion speed \cite{calder}. Finally an equilibrium is reached when the remaining pressure is dominated by gas that permeates from the outside through the wall of the vacuum vessel. Naturally the permeation rate is very small in typical situations, and diffusion coefficients in metals are relevant only for light molecules like hydrogen.  

So far we have considered thermal outgassing of materials as the main source of gas load in a~vacuum recipient. In the presence of an intense particle beam the release of gas molecules from the~chamber walls can be drastically enhanced, by orders of magnitude, and those effects must be taken into account and quantified. A prominent effect is photo-desorption due to synchrotron radiation (SR), emitted by an electron or positron beam \cite{fischer}. When a photon is absorbed on a metallic surface, the~photo effect may lead to the emission of a free electron. Absorption of the electron may then result in the~release of a gas molecule that was bound on the surface of the vacuum chamber. In a dipole magnet with bending radius $\rho$ the number of photons emitted per unit length and time is given by:

\begin{equation}
\frac {dN_\gamma}{dtds} = 1.28 \cdot 10^{17} \frac{I\,[\RM{mA}]\,E\,[\RM{GeV}]}{\rho\,[\RM{m}]}\,.
\end{equation}
Here $E$ is the beam energy and $I$ the beam current. The photons exhibit a statistical distribution of their energies and the given formula delivers the total number. More details on the properties of synchrotron radiation are given for example in Ref. \cite{vacuumelectronic}. The SR induced desorption results in a specific outgassing rate per unit length of:

\begin{equation}
q = \eta \, k_b\, T\, \frac {dN_\gamma}{dt\,ds}\,.
\end{equation}

\begin{figure} 
\centering\includegraphics[width=0.85\textwidth]{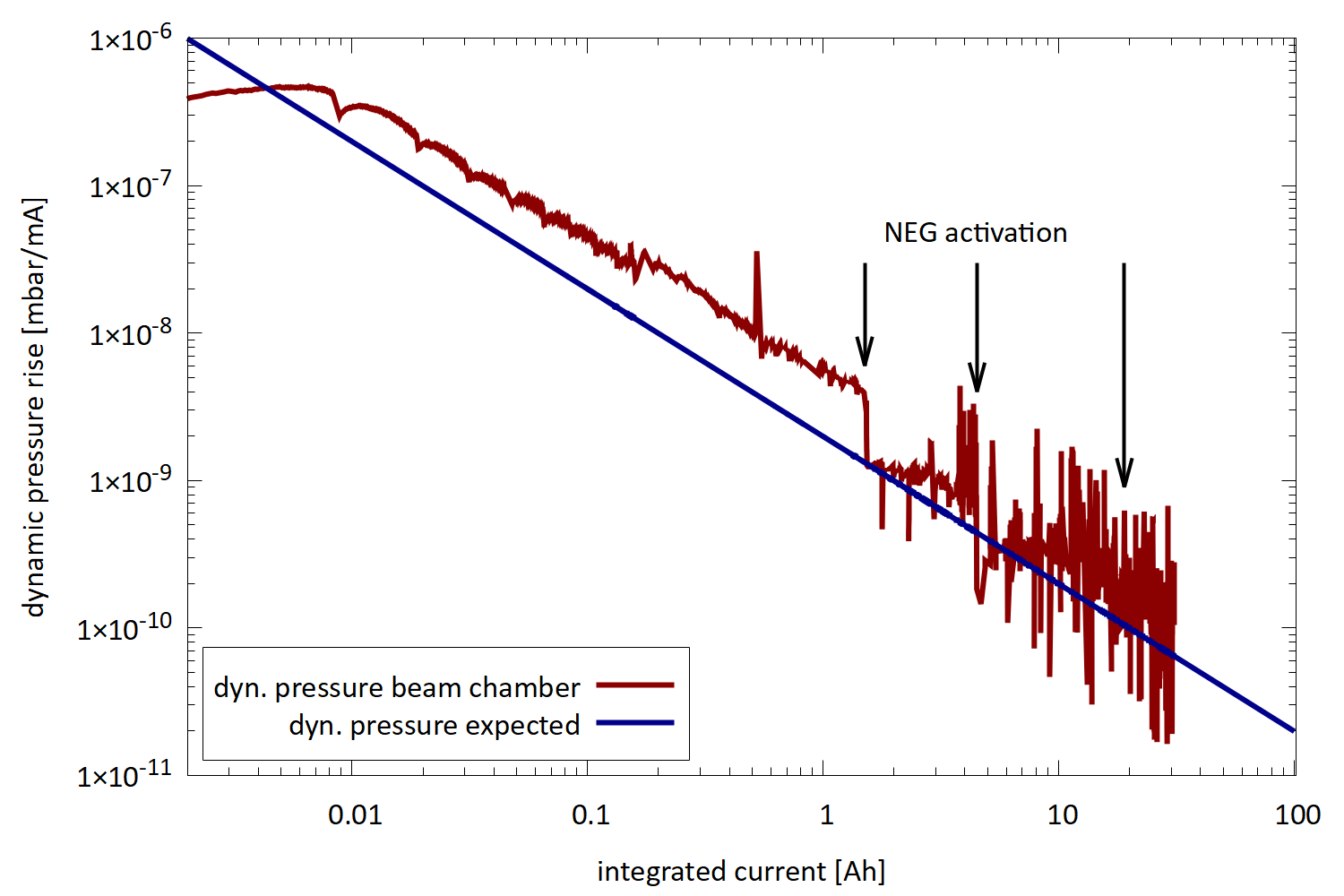}
\caption{The pressure rise per beam current is called dynamic pressure. It is shown here for a test chamber of the~PETRA-III storage ring as a function of integrated beam current.}
\label{fig:petra}
\end{figure}

The desorption yield $\eta$ equals the number of gas molecules released per absorbed photon. It depends on the material of the chamber wall, the preparation of the material and foremost the conditioning of the surface under the bombardment with photons. For a surface cleaned with standard procedures one observes a decrease of the desorption yield inversely proportional to the total number of absorbed photons, e.g. Ref.\,\cite{billy}. The pressure obtained in a storage ring as a balance between beam induced desorption and installed pumping speed is called dynamic pressure as it depends on the beam intensity. In Fig.\,\ref{fig:petra} the~conditioning process of a NEG pumped vacuum chamber for the synchrotron light source PETRA-III is shown. The pressure rise per beam current is reduced by several orders of magnitude over the course of the conditioning process. In a well conditioned vacuum system the desorption yield $\eta$ may reach values of $10^{-6}$ or lower. Rings with high intensity beams exhibit a high photon flux and one might be tempted to expect also high dynamic pressure for those. However, in this situation also the conditioning process advances faster and so the achieved pressure after a certain operating time is often similar for different storage rings. The typical gas composition is dominated by H$_2$ and a $\sfrac{1}{4}$ fraction of CO.

Besides the discussed mechanism of photo desorption there are other mechanisms causing beam induced desorption. If ion beams are accelerated, small losses of energetic ions may lead to high desorption rates. For example there are situations where a single ion, impinging on the wall of the vacuum chamber, may release of the order of $10^5$ gas molecules \cite{sen}. Ion induced desorption cannot easily be reduced by conditioning. 

Another important effect is the electron cloud instability \cite{ecloud}. A free electron is accelerated in the~electromagnetic field of the beam and releases more than one secondary electron when absorbed on the chamber wall. Due to the multiplicative behavior of the process, the intensity of these electrons grows exponentially, resulting in a proportional growth of gas release in the vacuum chamber. A key factor for this process is the secondary emission yield, that is the number of electrons released after the~absorption of one electron. It depends on the type of material and can be reduced by coating of the~surface, for example with TiN or C. Another countermeasure is the application of a small longitudinal magnetic field, by winding a coil around the beampipe, to force the low energy free electrons on spiral paths.

\section{Pressure Computation for One-Dimensional Systems}

Accelerator vacuum systems are build for beam transport and are typically lengthy, that is the longitudinal dimension is much larger than the transverse dimensions. For a calculation of the pressure profile a one-dimensional calculation is then sufficient. We start the derivation of the corresponding diffusion equation from the previously discussed relation between gas flow $Q$ and conductance $C$ in Eq. (\ref{eq:cond}). Note that we have to introduce at this point a minus sign that ensures positive gas flow towards smaller pressure. In the~limit of small differences we obtain a differential equation for the gas flow. Here a specific conductance $\mathcal{C} = C \Delta s$ is introduced, which is a property of the vacuum vessel cross section. 

\begin{eqnarray}
Q & \propto & - \d{P(s)}{s} \nonumber \\
Q(s) & = & -\mathcal{C}\cdot \d{P(s)}{s} 
\label{eq:flow}
\end{eqnarray}

A second equation is the continuity equation for the gas flow, which expresses that the change in the amount of transported gas is given by the sum of outgassing from the wall minus the pumped gas of a considered section with infinitesimal length. Here we use the specific pumping speed $\mathcal{S} = S/\Delta s$ and the specific outgassing rate $q$ per unit length in mbar\,l/m\,s.

\begin{equation}
\d{Q(s)}{s} = q - \mathcal{S}\,P(s)
\label{eq:continuity}
\end{equation}

These two Eqs. (\ref{eq:flow}) and (\ref{eq:continuity}) can now be combined into a~second order diffusion equation for a~one-dimensional vacuum system with the independent coordinate $s$.

\begin{equation}
\d{}{s}\,\mathcal{C}\,\d{}{s}\,P(s)-\mathcal{S}P(s) + q = 0
\label{eq:diffeq}
\end{equation}

Depending on the nature of the pumps two different types of solutions are obtained for Eq. (\ref{eq:diffeq}). In the most common situation of systems with lumped pumps the pumping speed $\mathcal{S}$ is non-zero only for short sections of the system. In-between these pumps the pressure distribution follows a quadratic function with a maximum half-way between a pair of pumps that are installed at a distance $l$:

\begin{equation}
P(s) = \frac{ql}{S}+ \frac{q}{8\mathcal{C}} \left(l^2-4s^2\right).
\label{eq:qprofile}
\end{equation}

Peak pressure and average pressure for this situation are given by:

\begin{equation}
P_\RM{avg} = ql\,\left(\frac{1}{S}+\frac{l}{12\,\mathcal{C}}\right), \quad
P_\RM{max} = ql\,\left(\frac{1}{S}+\frac{l}{8\,\mathcal{C}}\right).
\label{eq:aprofile}
\end{equation}

\begin{figure} 
\centering\includegraphics[width=0.925\textwidth]{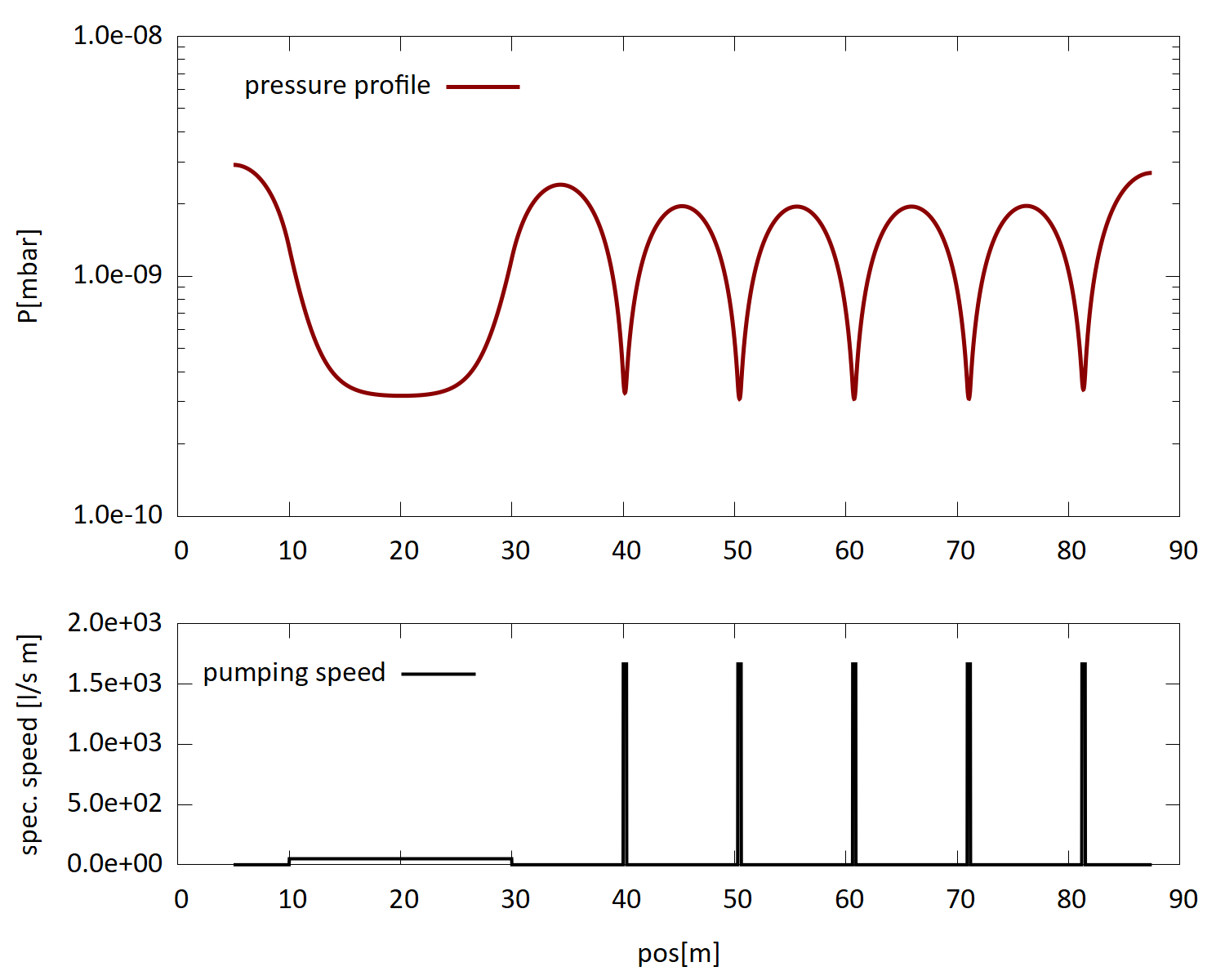}
\caption{Simulated pressure profile in the upper part of the image. The lower part shows the pumping speed along the beamline. A $d=10\,$cm beam pipe is assumed for this demonstration. While the left section contains a 20m long distributed pump, the  right section has 5 lumped pumps installed.}
\label{fig:v2}
\end{figure}

These functions have two terms. Even with infinite pumping speed (left term) the pressure is still limited by the conduction of the vacuum vessel. The economy of a technical solution is thus to be optimized w.r.t. density and size of the vacuum pumps, and the transverse dimensions of the vacuum vessel that determine the specific conductance.
Besides lumped pumps, also distributed pumps are used in accelerator vacuum systems. Distributed pumps can be realised by NEG strips or NEG coating, and by distributed ion sputter pumps that utilize the magnetic field of accelerator magnets. For the case of non-zero pumping speed we obtain an exponential solution of Eq. (\ref{eq:diffeq}). At some distance from neighboured sections the pressure in a long section with distributed pumps approaches a value of $P=q/\mathcal{S}$. Figure\,\ref{fig:v2} shows an exemplary calculation of a pressure distribution with lumped pumps in the right part and a long distributed pump in the left part of the graph.

Pressure profiles can also be calculated in a numerical approach using matrix multiplications \cite{ziemann}: 

\begin{eqnarray}
\left(\begin{array}{c} P(l) \\ Q(l) \end{array} \right) & = &
\left( \begin{array}{cc}
\cosh(\alpha l) & -\frac{1}{c\alpha} \sinh(\alpha l) \\
-\alpha c \sinh( \alpha l) & \cosh(\alpha l) \end{array} \right)
\left( \begin{array}{c} P(0) \\ Q(0) \end{array} \right)
+ \frac{q}{\alpha} \left(
\begin{array}{c} \frac{1-\cosh(\alpha l)}{\alpha c} \\
\sinh(\alpha l)  \end{array} \right). \hphantom{8888} 
\label{eq:matrix}
\end{eqnarray}

Here the variable $\alpha = \sqrt{\mathcal{S}/\mathcal{C}}$ was introduced. For sections without pumping, $\mathcal{S}=0$, the elements containing $\alpha$ in Eq. (\ref{eq:matrix}) have to be taken in the limit $\alpha\rightarrow 0$, leading to the quadratic profile Eq. (\ref{eq:qprofile}). So far we have considered the static situation of equilibrium pressure profiles. This is a special case of the general time dependent diffusion equation: 

\begin{equation}
\mathcal{V} \d{}{t} P(s,t) = \d{}{s}\,\mathcal{C}\,\d{}{s}\,P(s,t)
-\mathcal{S}P(s,t) + q.
\label{eq:timediff}
\end{equation}

The specific volume $\mathcal{V}$ denotes the volume of the vessel per unit length. This equation describes also dynamic evolutions of the pressure profile, for example when He gas is injected for reasons of leak search. By comparison with a classical diffusion equation $\d{}{t} f(x,t) = \d{}{x}\,\mathcal{D}\,\d{}{x} f(x,t)$ that is known from the literature, we identify the~diffusion coefficient of our problem as:

\begin{equation}
\mathcal{D} = \frac{<\Delta x^2>}{<\Delta t>} = \frac{\mathcal{C}}{\mathcal{V}}.
\label{eq:diffcoeff}
\end{equation}

\begin{figure} 
\centering\includegraphics[width=0.85\textwidth]{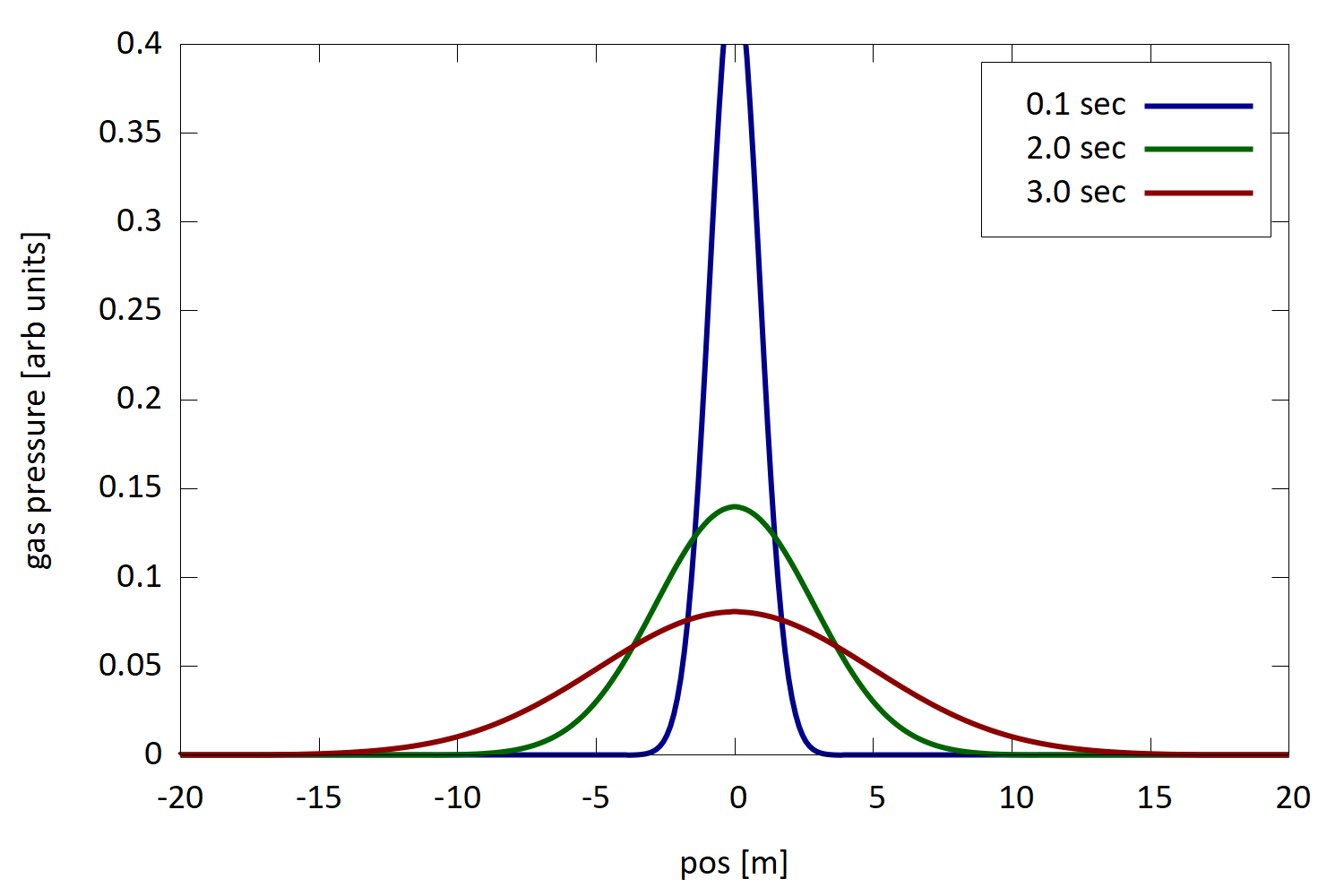}
\caption{A pressure bump of He in this example diffuses in a 2\,cm diameter tube at limited speed.}
\label{fig:diff_He}
\end{figure}

Using this parameter one can estimate that it takes 3 seconds for He gas to travel a 5\,m distance in a 2\,cm pipe. Figure\,\ref{fig:diff_He} shows the time evolution of the pressure distribution for a Delta-function like inlet of He gas in such a pipe.

The shown analytic calculations allow one to estimate the pressure in vacuum systems and to calculate the required number of pumps and their distance. For complex geometries Monte Carlo methods for pressure calculations might be more accurate. The code MolFlow \cite{molflow} follows individual gas molecules on their scattered path through a vacuum system, including sticking probabilities and sojourn times. Pressure and other variables are then calculated as statistical averages. For lepton storage rings the~code may be augmented by another code, SynRad \cite{synrad}, which allows to simulated the photo desorption process.

\section{Requirements for Accelerator Vacuum Quality}

In order to assess the required vacuum quality in terms of pressure or density it is necessary to study the~different mechanisms of beam gas interaction. Scattering of beam particles may lead to immediate loss of beam particles, or to a degradation of the beam properties. The number of lost particles $\Delta N_b$ from a beam passing through a section of length $\Delta l$ can be estimated in the following way:

\begin{eqnarray}
\Delta N_b & = & -N_b \times \frac{\mathrm{area\,of\,molecules}}{\mathrm{total\,area}} \nonumber \\
& = & -N_b \times \frac{n_v V \sigma}{V/\Delta l} \nonumber \\
& = & -N_b n_v \sigma \Delta l \nonumber \\
& = & -N_b n_v \sigma \beta c \Delta t \, . \nonumber
\label{eq:cross}
\end{eqnarray}

Here $\sigma$ is the cross section of a generic interaction process, $N_b$ is the number of beam particles and $n_v$ is the volume density of the residual gas. In case the considered scattering mechanism leads to the loss of the particle, and by converting this relation into a differential equation, the solution is given by an exponential decay of the beam intensity:

\begin{equation}
N_b(t) = N_0 \exp \left( - \sigma \beta c n_v \, t \right), 
~\tau = \frac{1}{\beta c \, \sigma \, n_v}.
\label{eq:diffcoeff}
\end{equation}

The beam lifetime $\tau$ is deduced from the exponential solution, and for most situations we can safely assume $\beta=1$. In the following we consider the effect of different scattering mechanisms for electrons and protons.\hfill\\

\begin{figure} 
\centering\includegraphics[width=0.85\textwidth]{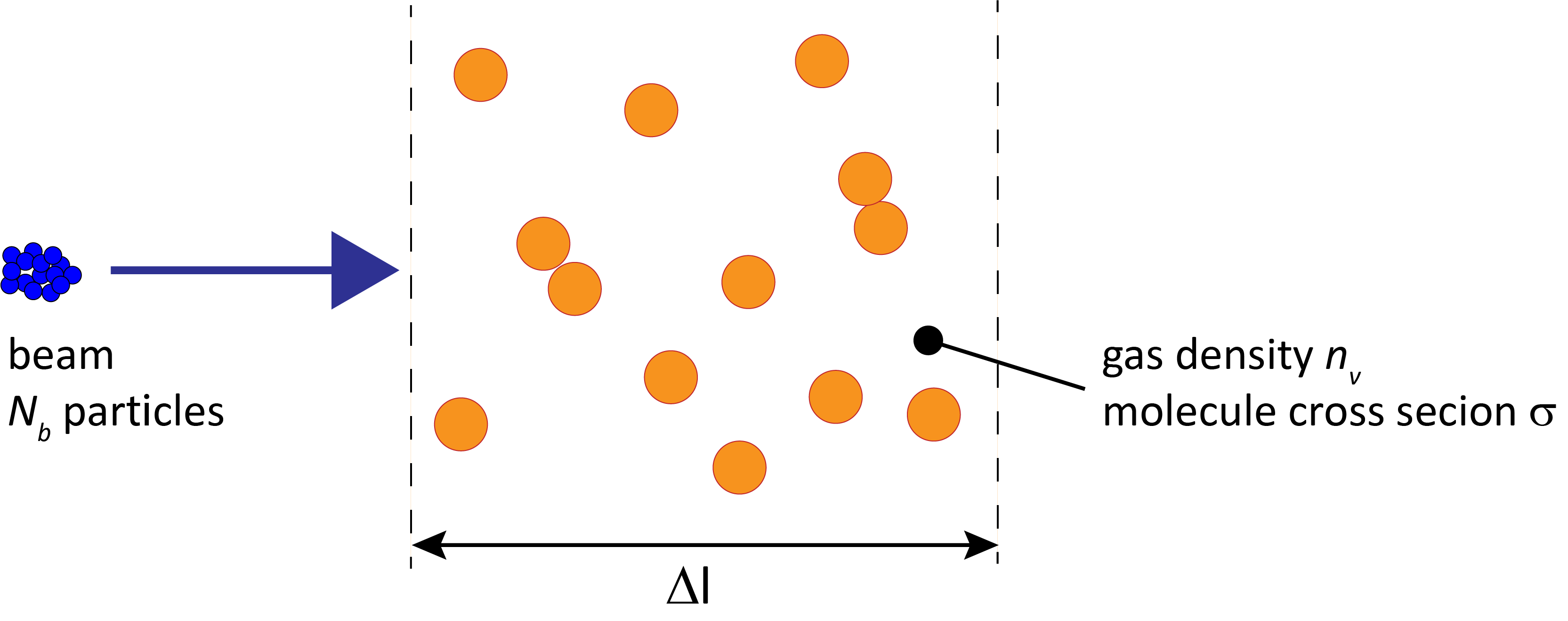}
\caption{A beam with $N_b$ particles passes through a volume with residual gas density $n_v$. The gas molecules exhibit an effective cross section $\sigma$ that describes the interaction probability with the beam for specific processes.}
\label{fig:beam_gas}
\end{figure}

\subsection{Coulomb Scattering for Electrons}

\noindent
This process is described by the well known formula for Rutherford Scattering, which gives the differential cross section for the occurrence of a scattering angle $\theta$:

\begin{equation}
\frac{d\sigma_i}{d\Omega} = \frac{Z_i^2\,r_e^2}{4\gamma^2}~
\frac{1}{\sin^4(\theta/2)}.
\label{eq:rutherford}
\end{equation}

The charge of the residual gas atom is $Z_i$ and $r_e=2.8\,$fm the classical electron radius. If we integrate this differential cross section from the angle $\theta_0$ above which particles are lost to $\pi$ and use $\theta_0 \ll 1$, the total elastic scattering cross section for particle loss is:

\begin{equation}
\sigma_{i, \RM{el}} = \frac{2\pi\,Z^2_i\,r_e^2}{\gamma^2}~
\frac{1}{\theta_0^2}.
\label{eq:elastic}
\end{equation}

The limiting angle $\theta_0$ can be estimated from a typical value for the $\beta$-function $\overline{\beta_y}$ and the minimum aperture $A_y$ of the accelerator: $\theta_0 = A_y/\overline{\beta_y}$. Using the parameters for the vertical plane is sufficient since the vertical aperture is usually smaller than the horizontal one in electron storage rings. Average values of the $\beta$ functions at the locations of particle scattering and particle loss have to be used. Combining the~above relations and carrying out the sum over different atom species we obtain the following formula for the beam lifetime due to elastic scattering:

\begin{equation}
\tau^{-1}_\RM{el} = \frac{2\pi r_e^2 c}{\gamma^2}~
\frac{\overline{\beta_y}^2}{A_y^2} \sum_i n_i \sum_j k_{ij} Z_j^2.
\label{eq:telastic}
\end{equation}

Here $k_{ij}$ is the number of atoms of type $j$ within the molecule of type $i$. By inserting numbers for the fundamental constants and expressing the gas density in terms of pressure at room temperature we obtain the following formula for the beam lifetime in electron rings due to elastic scattering:

\begin{equation}
\tau_\RM{el}\,[\RM{h}] = 2839~\frac{E^2\,[\RM{GeV}^2]\,\,A_y^2\,[\RM{mm}^2]}
{\overline{\beta_y}^2\,[\RM{m}^2]}~
\left( \sum_i P_i\,[\RM{pbar}] \sum_j k_{ij} Z^2_j \right)^{-1}.
\label{eq:handy}
\end{equation}

Note that the quadratic function of $Z$ causes a sensitive dependence of the beam lifetime on the~presence of heavy gas species in the gas composition.\hfill\\

\subsection{Bremsstrahlung} \noindent
Due to deceleration of a beam particle in the Coulomb field of a residual gas atom and the emission of a high energy photon, the particle may leave the energy acceptance of the accelerator. The important parameter in this context is the largest allowed relative energy deviation for the particles to stay confined within the beam: $\delta_E = \Delta E/E_0$.  The cross section for the inelastic process is

\begin{equation}
\sigma_\RM{inel} \approx -\frac{4}{3}\,\, \frac{V_n}{N_A}\,\,\frac{1}{X_0}\,\ln\,\delta_E.
\label{eq:inelastic}
\end{equation}

From this the following lifetime $\tau_\RM{brems}$ is computed. For a gas mixture one has to sum up contributions of gas species with their partial pressures $P_i$ and corresponding radiation lengths $X_{0,i}$. 

\begin{eqnarray}
\frac{1}{\tau_\RM{brems}} & = & -\frac{4}{3}~\frac{c}{P_n}\ln(\delta_E)~ \sum_i \frac{P_i}{X_{0,i}} \nonumber \\
\tau_\RM{brems}\,[\RM{h}] & = & \frac{-0.695}{\ln(\delta_E)}\,\left(\sum_i \frac{P_i\,[\RM{pbar}]}{X_{0, i}\,[\RM{m}]}\right)^{-1}.
\label{eq:brems}
\end{eqnarray}

$N_A$ is the Avogadro constant and $V_n=22.4\,$l/mol, $P_n$ the molar volume and the pressure under standard conditions. The \emph{radiation length} $X_0$ is the length over which a particles energy has dropped by a factor $1/e$. $X_0$ scales roughly inversely proportional to the square of the nuclear charge of the~residual gas, and also inversely proportional to its density.  Radiation length values for common gases under normal conditions are tabulated for example in Ref. \cite{pd}. In Table \ref{tab:radlength} we list $X_0$ for the important gases of accelerator vacuum systems. With common energy acceptance and transverse acceptance of storage rings, the mechanism of Bremsstrahlung is the more severe mechanism for loss of particles from the~beam.

\subsection{Emittance Growth for Hadrons} \noindent
For a proton or ion beam already the degradation of the beam emittance from elastic gas scattering at small angles is harmful. Due to the absence of radiation damping any decrease of the beam density over the storage time cannot be recovered. For the beam emittance a growth time can be defined:

\begin{equation}
\frac{1}{\tau_\varepsilon} = \frac{1}{\varepsilon_x}~\frac{d\varepsilon_x} {dt}.
\label{eq:emmlife}
\end{equation}

The scattering causes a diffusive growth of the mean squared angular deviation of the particles momentum vector which is linear in time.  The emittance growth is related to this angle as follows ($\theta_0$ is the rms scattering angle projected on a transverse plane):

\begin{equation}
\frac{d\varepsilon}{dt} = \frac{1}{2}~\overline{\beta_y}~\frac{d(\theta_0^2)}{dt} = 
\frac{1}{2}~\overline{\beta_y}~\frac{(13.6)^2}{(cp)^2\,[\RM{MeV}^2]}~\frac{c}{P_0}~
\sum_i \frac{P_i}{X_{0,i}}.
\label{eq:emmgrowth}
\end{equation}

Using Eq. (\ref{eq:emmlife}) the resulting emittance growth time for protons is:

\begin{equation}
\tau_\varepsilon\,[\RM{h}] \approx 34.2~\frac{\varepsilon_y\,[\RM{m\,rad}]\,
E^2\,[\RM{GeV}^2]\,T\,[\RM{K}]}{\overline{\beta_y}\,[\RM{m}]}~\left(
\sum_i \frac{P_i\,[\RM{pbar}]}{X_{0, i}\,[\RM{m}]}\right)^{-1}.
\end{equation}

The temperature $T$ has been included since proton accelerators often use superconducting magnets and cold beam pipes. The described mechanism ignores other elastic scatting mechanisms besides Coulomb scattering. \hfill\\

\subsection{Inelastic Scattering for Hadrons}

Another process is the complete removal of particles from the beam by an inelastic reaction. The beam lifetime for this effect can be computed using the inelastic interaction length $\lambda_\RM{inel}$ which is also tabulated in Table\,\ref{tab:radlength}:

\begin{equation}
\frac{1}{\tau_\RM{inel}} = \frac{\beta c}{P_0}~
\sum_i \frac{P_i}{\lambda_{\RM{inel},i}}.
\label{eq:tauinel}
\end{equation}

The inelastic interaction length is related to the corresponding nuclear cross section via $\lambda_\RM{inel} = A/\rho\,N_A\,\sigma_\RM{inel}$, where $A$ is the molar mass and $\rho$ the density. If we again include the gas temperature and take out all constants we obtain the following formula for the inelastic beam-gas lifetime:

\begin{equation}
\tau_\RM{inel}\,[\RM{h}] = 3.2\cdot10^{-3}~T\,[\RM{K}]\,\left(
\sum_i \frac{P_i\,[\RM{pbar}]}{\lambda_{\RM{inel}, i}\,[\RM{m}]}\right)^{-1}.
\end{equation}

\begin{table}
\begin{center}
\caption{Radiation length $X_0$ and inelastic interaction length $\lambda_i$ 
for different gases under atmospheric pressure and 20\,$^\circ$C \cite{pd}.}
\label{tab:radlength}
\begin{tabular}{|l|cc|cc|cc|cc|cc|cc|cc|cc|cc|c|}
\hline\hline
&& \textbf{H}$_\textbf{2}$ && \textbf{He} && \textbf{CH}$_\textbf{4}$ && \textbf{H}$_\textbf{2}$\textbf{O} &&\textbf{CO}&&\textbf{N}$_\textbf{2}$&&\textbf{Ar}&&\textbf{CO}$_\textbf{2}$&&\textbf{air} \\
\hline
A && 2   && 4  && 16     && 18     &&28&& 28  &&40&&44    &&     \\
$X_0$\,[m] && 7530 && 5670 && 696 && 477 && 321&&326&&117&&196&& 304    \\
$\lambda_\RM{inel}$\,[m]&&6107&&3912&&1103&&1116&&763&&753&&704&&490&&747 \\
\hline\hline
\end{tabular} \end{center}
\end{table}

\section{Vacuum Technology for Accelerators}
\subsection{Pumping}
\label{pumping}

\begin{figure} 
\centerline{\includegraphics[width=0.9\textwidth]{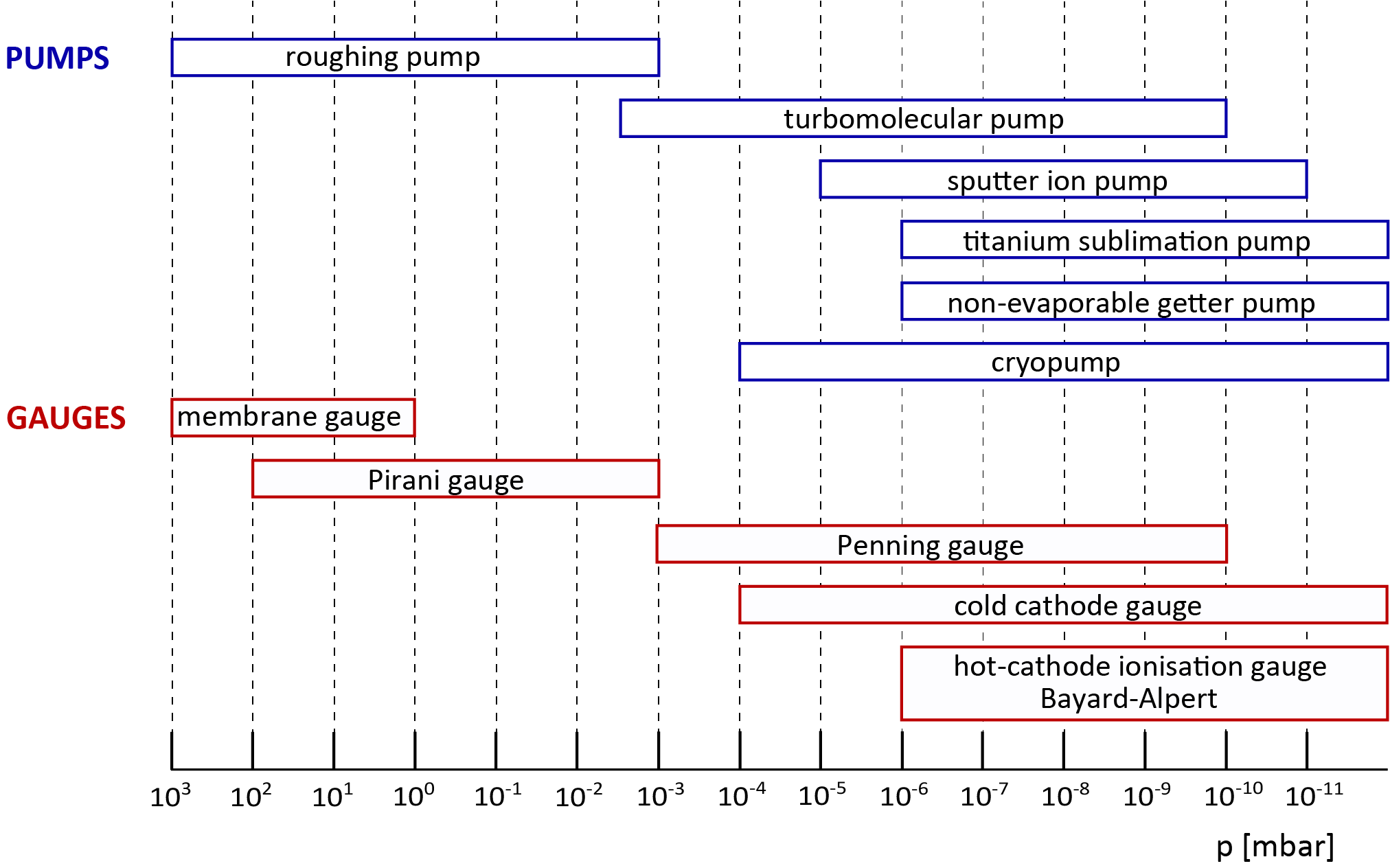}}
\caption{Overview of the most commonly used types of pumps and gauges in large accelerator vacuum systems including their operating range.} 
\label{fig:pumps}
\end{figure}

As discussed in the previous sections the average pressures required for accelerators range from $10^{-6}\,$mbar down to $10^{-11}\,$mbar. For facilities of large size, containing a huge number of components, reliability of the overall system becomes a critical aspect. Pumps without moving mechanical parts are thus advantageous. Figure\,\ref{fig:pumps} shows an overview of the most commonly used types of pumps in large accelerator systems. Initial pumpdown of accelerator vacuum systems is often done by turbo-molecular pumps in combination with rotary pumps that are installed on mobile pump carts. During normal operation these are disconnected while the required vacuum conditions are maintained by sputter ion pumps, titanium sublimation pumps (TSP) and non-evaporable getter (NEG) pumps. These types of pumps are almost exclusively used for routine operation in storage rings and large linear accelerators. 

\emph{Turbo-molecular pumps} are fully mechanical pumps, based on a mechanism to transfer physical momentum to residual gas molecules in a preferred direction. The momentum transfer is achieved through fast moving blades with rotation frequencies of $30\dots60\cdot 10^3\,$ RPM (rounds per minute). At these rotation frequencies the blades reach a speed of $300\dots600\,$m/s, which is to be compared with the speed of the residual gas molecules that should ideally not escape the blades. At room temperature heavier molecules like CO have a speed of $470\,$m/s, while hydrogen moves at $1800\,$m/s. Consequently the compression ratio (gas density ratio) achieved with a turbo-molecular pump may vary by several orders of magnitude for these gas species. Turbo-molecular pumps are always combined with mechanical roughing pumps that provide an intermittent vacuum level between UHV conditions and the normal atmospheric pressure.

\emph{Sputter ion pumps} are capable of pumping all gases and they can be operated at relatively high pressure. These pumps use penning cells with an applied high voltage of 3\dots7\,kV and a superimposed magnetic field. Permanent magnet blocks are used to generate the magnetic field. Residual gas molecules are ionized and accelerated in the electric field. The current drawn from the high voltage power supply is thus proportional to the residual gas pressure. This synergy is often used to measure the pressure in accelerators in a cost effective way. Hydrogen is pumped by diffusion into the bulk of the cathodes. All other reactive gases are chemisorbed by the cathode material. A commonly used material is titanium, which is sputtered onto the walls and anodes by the ions upon incident on the cathodes. Noble gases are physically buried by the sputtered cathode atoms. The pumping speed for noble gases is small, but it can be increased to values of $25-30\,$\% of that for N$_2$ replacing the cathode plate by heavy material such as tantalum. Such \emph{noble diodes} are often used in systems with enhanced risk of helium leaks as in accelerator sections 
using superconducting magnets or resonators cooled with liquid helium.

\begin{figure} 
\centering\includegraphics[width=0.75\textwidth]{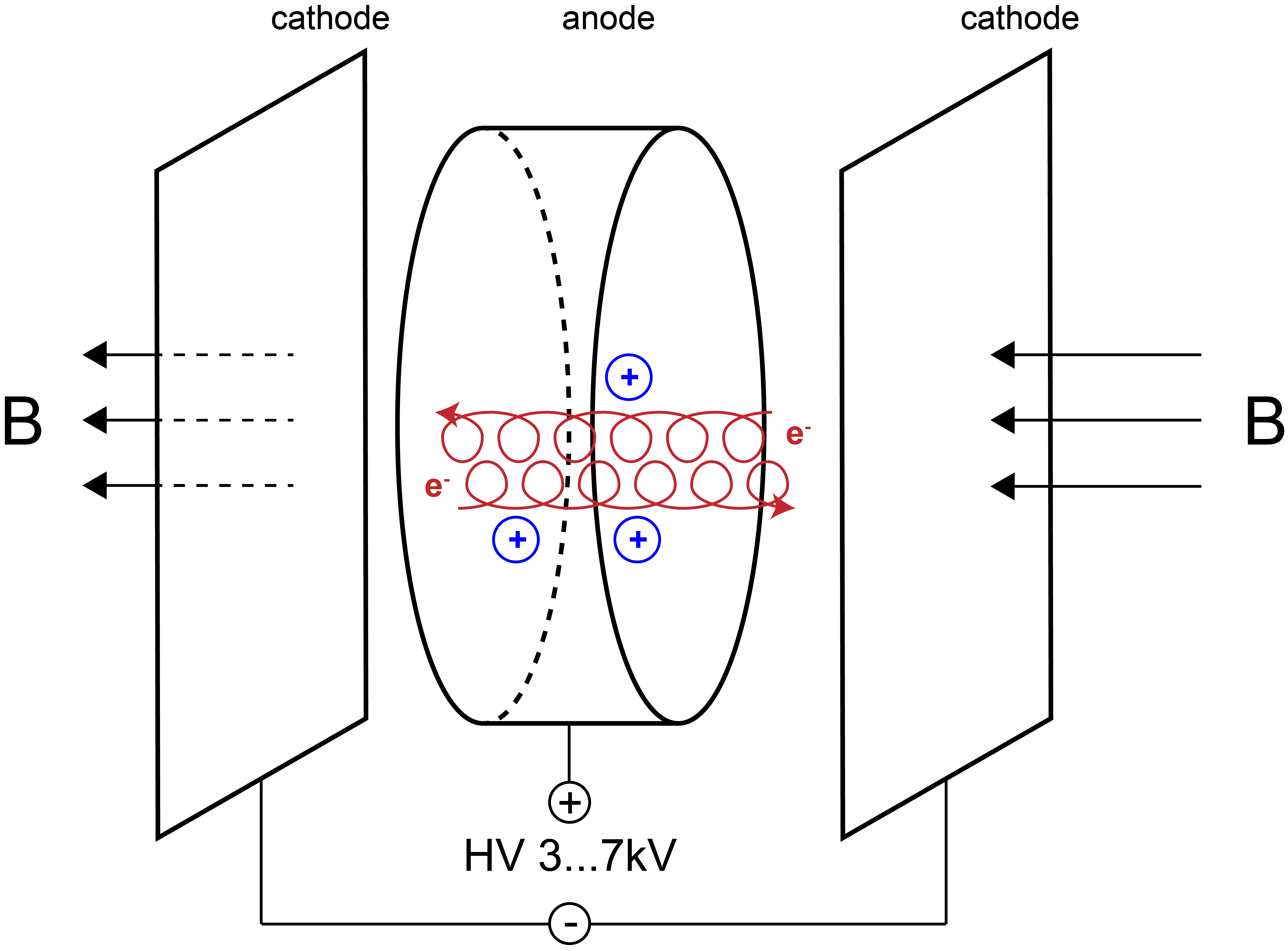}
\caption{Penning cells are elements in ion sputter pumps to ionize residual gas molecules. In a combined electric and magnetic field electrons are accelerated on spiral paths, thereby enhancing their efficiency to ionize neutral gas molecules. Ions are then also accelerated and implanted in the cathod material or burried under sputtered metal on the anodes.}
\label{fig:penning}
\end{figure}

A cost effective solution are so-called integrated sputter ion pumps inserted linearly into a channel parallel to the beam channel \cite{cummings}, \cite{koupt}. These pumps utilize the magnetic field of the bending magnets of the accelerator. This solution has been adopted at the electron ring of HERA, PETRA, PEP and TRISTAN reaching typical pump speeds of $25\,$l/s/m. On the downside this couples the functions of the accelerator magnets and the vacuum system. Magnets must be powered to maintain pumping. In-between ramp cycles of storage rings these pumps are not active. Furthermore the magnetic field intensity is decreased when the particles are injected at lower energies, resulting in a reduced pumping speed. This becomes crucial for accelerators where the injection energy and thus the corresponding magnetic field are much lower than the nominal beam energy, such that the discharge in the sputter ion pumps extinguishes. Thus additional pumps are required to ensure good vacuum conditions. 

\emph{Titanium sublimation pumps} are sorption pumps. In a metallic vessel titanium
is evaporated by temporary electrical heating and deposited on the walls, forming a getter surface. These pumps exhibit a~high pumping speed for active gases but have limited pumping capacity since the thin titanium film saturates quickly. The pumping surface is renewed
by deposition of fresh titanium from a heated filament by sublimation. A pump with $1000\,$l/s pumping speed is saturated after one hour at a pressure of $10^{-7}\,$mbar. Titanium sublimation pumps cannot pump noble gases and are therefore used in combination with a low pumping speed ion sputter pump. Often a chicane is incorporated in the pumping port, to avoid that titanium atoms enter the beam chamber or contaminate surfaces, e.g. mirrors, ceramic insulators or instrumentation.  

\begin{figure} 
\centering\includegraphics[width=0.85\textwidth]{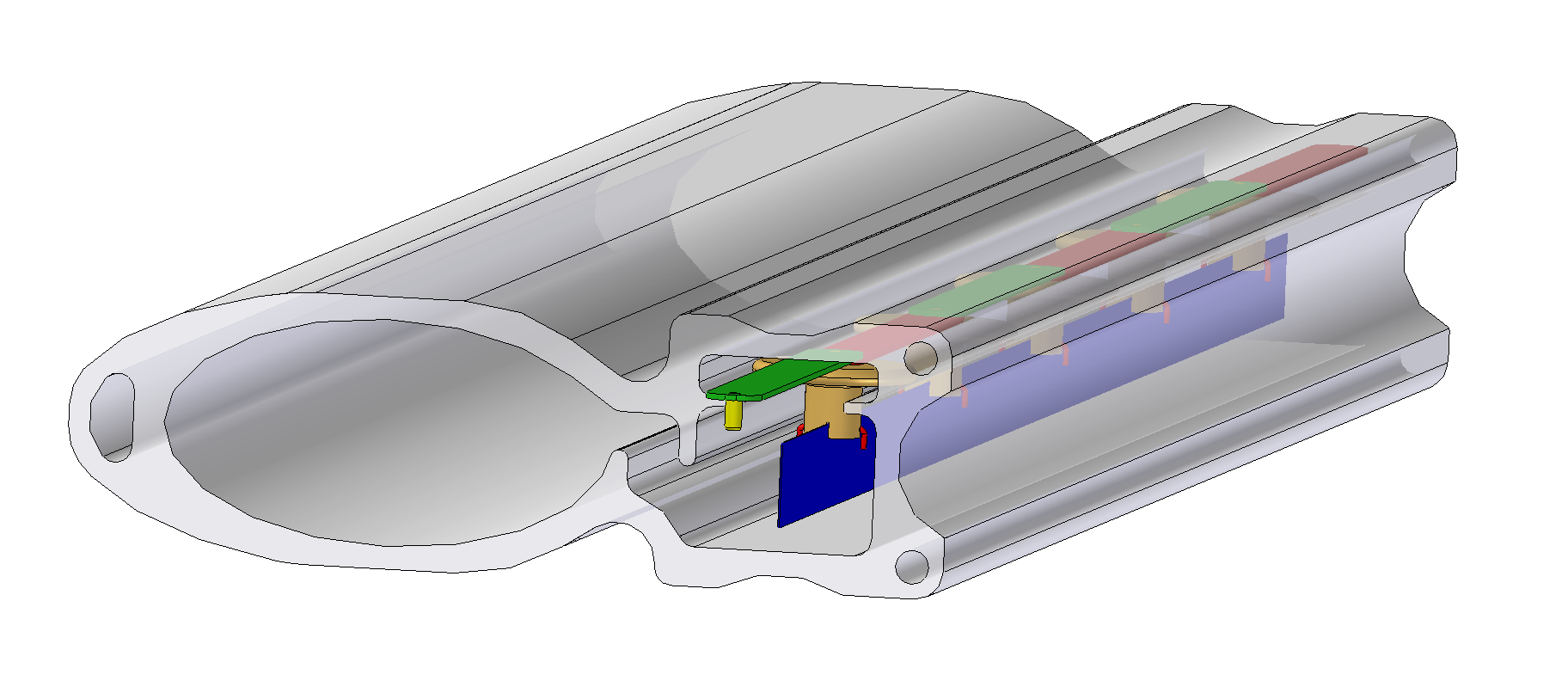}
\caption{Concept view of a vacuum chamber with integrated NEG pump for an electron beam storage ring. The~aluminum profile is extruded with integrated cooling channels. The synchrotron radiation is absorbed on the left side of the chamber, while the right side contains the NEG strip (blue) in a side channel, which is shielded from direct view to the beam to suppress capture of dust particles (courtesy DESY).}
\label{fig:chamber}
\end{figure}

\emph{Non-evaporable getter (NEG) pumps} are sorption pumps as well. The NEG material is made of special alloys which forms stable chemical compounds with the majority of active gas molecules. The~sorption of hydrogen is reversible by heating of the material. Also NEG pumps have a limited pumping capacity. The NEG material is activated by heating for times below one hour. Activation temperatures depend on the NEG material and range from $180^\circ$C to $400^\circ$C. During heating the gas molecules are not evaporated from the NEG material, but the molecules diffuse into the bulk material. Hydrogen is an exception, which is released again into the gas phase, thus requiring other pumps during the process of activation and reactivation. The heating produces fresh surface sites for further adsorption of active gases. The NEG material is a compound of different metals, for example Zircon, Vanadium, Iron, and is typically sintered in the form of a powder onto flexible strips. Such strips can be integrated in a side channel of the vacuum chamber design. The initial specific pumping speed provided by such schemes may exceed 1000\,l\,s$^{-1}$\,m$^{-1}$. This technique has been developed for LEP \cite{ben-NEG} and is now applied in many accelerators. At LEP about 24 km of the beam pipe have been equipped with 30\,mm wide and 0.2\,mm thick constantan ribbons coated with 0.1 mm thin NEG material on both sides. A newer application for the synchrotron light source PETRA-III is shown in Fig.\,\ref{fig:chamber}. The strips are installed onto a~rigid stainless steel carriage via insulating ceramics inside a separate pump channel. For electric heating the pumps are connected to current feedthroughs. NEG pumps are also available commercially as lumped pumps, cartridge units that can be connected to recipients by standard flange connections.

Another approach to the NEG pumping concept is the deposition of thin film coatings of TiZrV, sputtered onto a vacuum chamber. The activation temperature of these coatings is relatively low, a fact that is important to limit the thermal stress on the vacuum system during activation \cite{Benvenuti-coating-1}, \cite{Benvenuti-coating-2}. If the~design of the vacuum chamber allows that, the coating may cover the complete inner surface of the~beam vacuum system and thus the outgassing of the vacuum vessel itself is drastically reduced. It should be emphasised that even with low activation temperature the need for baking the entire vacuum system to ca. $200\,^\circ$C implies severe restrictions for the design of vacuum chambers, support structures and magnets. For accelerators in which the vacuum conditions are dominated by beam induced desorption, it is possible to reduce outgassing with NEG coated surfaces in the vicinity of the particle beam. In particular for light sources NEG coated surfaces are advantageous to reach good vacuum conditions without long conditioning times \cite{chiggiato}. Modern light sources use complex multi-bend achromat lattice cells to achieve extremely small horizontal emittances. As a result the apertures of vacuum chambers are small, and the achievable pressure is conductance limited. Lumped pumping concepts are less efficient in this situation since a large number of pumps had to be installed per unit length. Today it is common for such light sources to use NEG coating for efficient pumping of the narrow beam chambers. Also hadron accelerators benefit from NEG coated chambers through the reduction of the secondary emission yield of electrons \cite{wfischer}.

\begin{figure} 
\centering\includegraphics[width=0.575\textwidth]{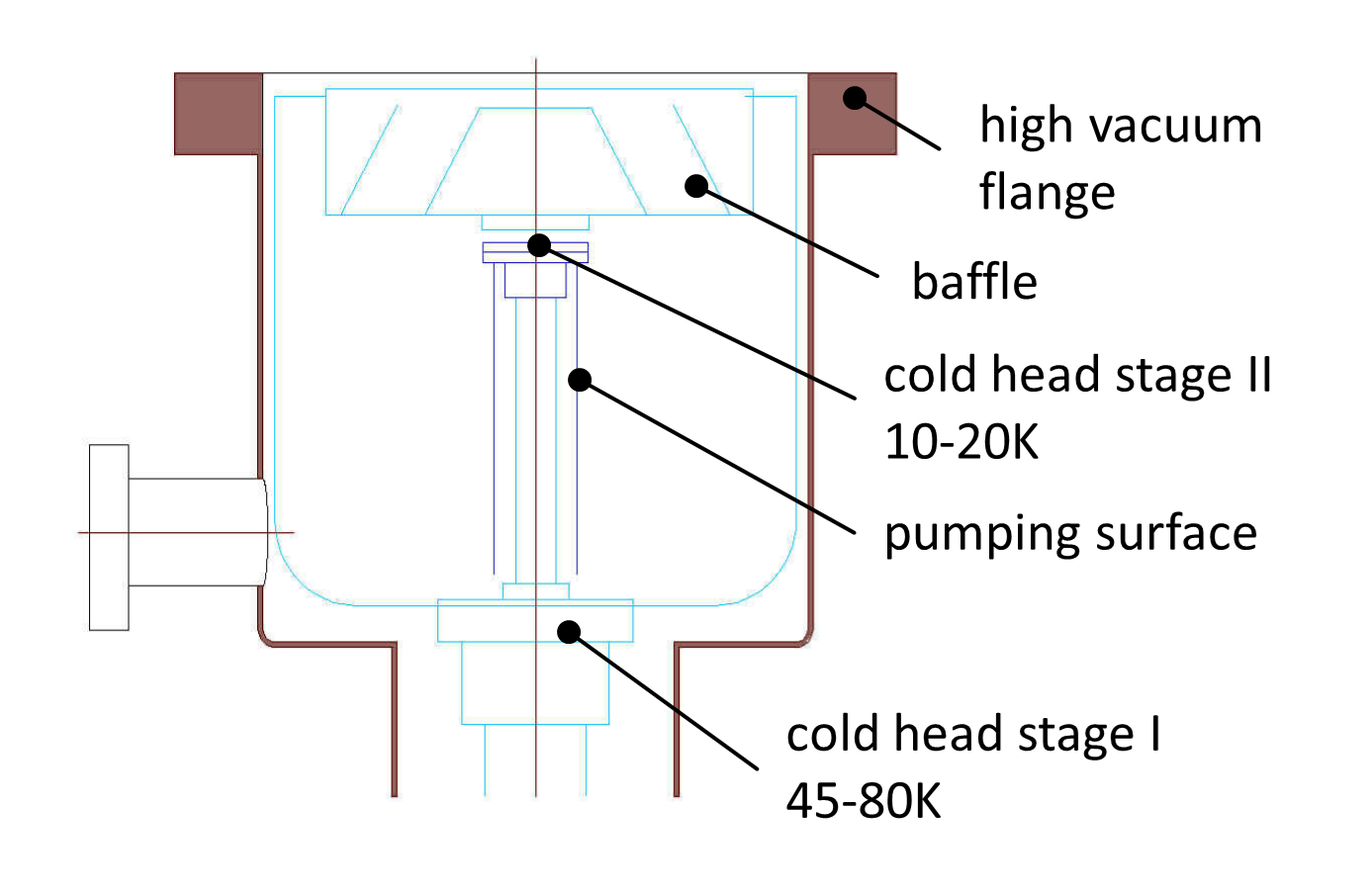}
\caption{Concept view of a cryo pump with a large aperture flange connection on the top side (courtesy Lothar Schulz, PSI).}
\label{fig:cryo}
\end{figure}

\emph{Cryo pumps} use the effect of cryosorption to bind gas molecules on cold surfaces inside the pump vessel. With sufficiently low temperature these pumps can remove all gas species and they provide high pumping speed and capacitance. On the downside cryo pumps must be regenerated by regular warm-up cycles, for which the accelerator operation must be interrupted. During the regeneration cycle all pumped gases are released again, and are typically removed from the vacuum system by turbo molecular pumps. A concept sketch of a cryo pump is shown in Fig.\,\ref{fig:cryo}.

For performance reasons many accelerators make use of superconducting magnets or superconducting accelerating structures. With the beam pipe being integrated into the cryostat this leads to a \emph{cold bore} vacuum system, which takes on the characteristics of a huge cryopump. Design and operation of such cold bore vacuum systems have many specific implications \cite{grobner}.

\subsection{Instrumentation} \label{instrumentation}

Accelerator based research infrastructures are expensive installations, they use a lot of grid energy and their operation should be efficient and reliable. Also for vacuum systems it is therefore important that the integrity of the system is continuously monitored and problems like leaks or regions of unusual high pressure resulting from beam impact can be identified quickly to allow fast and targeted intervention. Using gate valves the beam vacuum system of accelerators is divided into several sections. In this way it is possible to exchange components without venting the entire facility. Fast shutters, capable to stop shock waves in milliseconds, are commonly used to avoid contamination of sensitive sections caused by a sudden break of the vacuum system.

Each of the segments of a larger vacuum system should have at least one gauge to monitor the~integrated residual gas pressure. Total pressure gauges are included in Fig.\,\ref{fig:pumps} with their operating range. For practical operation it is usually not necessary to obtain precise absolute measurements, but to be able to diagnose relative changes over time and to compare sections with each other. Cold cathode gauges or Bayard-Alpert gauges are frequently used at low pressures. Current monitoring of sputter ion pumps is a~cost effective method, particularly for large facilities to monitor the residual gas pressure down to levels of $10^{-9}\,$mbar. 

Using quadrupole mass spectrometers the residual gas composition in a vacuum system can be analysed. The relative occurrence of molecule masses is determined, which allows to diagnose problems in an accelerator vacuum system. For example it can be determined whether a leak or a contamination is the reason for unusually high pressure. Such residual gas analysers (RGA) are typically not part of a standard installation, but these are temporarily connected to the recipient. Often the sensitive electronics is not compatible with the radiation environment of the stray magnetic fields of an accelerator in operation. 

\subsection{Materials and Technology Choices for Accelerator Vacuum Systems}

For large facilities production cost can be optimized by a careful design and by utilization of industrial manufacturing processes. For example extrusion processes serve as a cost effective method to produce large lengths of beam pipe profiles. Over time a wide variety of best practices have been developed for accelerator vacuum systems and more details may be found for example in Ref. \cite{CERN}.

Materials for the manufacturing of beam chambers and other beam vacuum components have to be selected carefully according to the specific requirements of each accelerator. A significant number of aspects must be addressed in parallel. The air pressure under normal conditions on a surface area of $10\times10\,\RM{cm}^2$ results in a force equivalent to the weight of 100\,kg. Consequently mechanical robustness is already one important criterion for a vacuum chamber, while the costly aperture of accelerator magnets might favour a thin wall solution. Vacuum chambers should be bake-able and thus stability must be guaranteed also at elevated temperatures. The magnetic guide field for the beam should not be disturbed by magnetic properties of the beam chamber. If synchrotron radiation is deposited on the chamber wall a high thermal conductivity is desirable. And the transport of beam image currents in the chamber walls requires good surface conductivity and smooth electrical connections across components. For machines susceptible to electron cloud instabilities the secondary electron emission yield of the material should be low. Last but not least the compatibility with UHV vacuum conditions and low outgassing rates are important. The requirement of UHV class pressure in combination with resistance to radiation and corrosive atmospheres demands all metal solutions for beam vacuum system.  

Stainless steel, copper or aluminum are the most common vacuum chamber materials. Electrical and thermal conductivity are by factors better for copper and aluminum surfaces compared to steel. On the other hand the mechanical strength of steel is outstanding. Copper and aluminum alloys exhibit better mechanical strength than the pure metals, at the expense of somewhat lower conductivity. A~few properties for common materials are listed in Table\,\ref{tab:materials}. Joining techniques for materials of vacuum systems include inert gas shielded arc welding, electron-beam welding, laser beam welding as well as brazing in a~furnace. Laser and electron beam welding can deposit large amounts of heating power to a~small volume in a short time. These technologies can be used advantageously for welding of sensitive components that allow only local heating while another part of the component must be kept at moderate temperature.

\begin{table}
\begin{center}
\caption{Properties of materials that are utilized for accelerator vacuum components.} 
\label{tab:materials}
\begin{tabular}{|l|cc|cc|cc|cc|} 
\hline\hline
                      && \textbf{density}    && \textbf{thermal}      && \textbf{electrical}         && \textbf{yield}      \\   
                      &&            && \textbf{conductivity} && \textbf{conductivity}       && \textbf{strength}   \\   
	                  && \textbf{[g/cm}$^3$] && \textbf{[W/K/m]}      && \textbf{[10}$^6/ \Omega$\textbf{/m]} && \textbf{[N/mm}$^2$] \\ \hline 
stainless steel 316LN && 8.00  	    && 16     	    && 1.35 		      && 205        \\ 
aluminum pure         && 2.70	    && 235     	    && 37  		          && 35         \\  
AlMgSi0.5             && 2.70       && 200          && 30  		          && 70-150     \\  
copper pure           && 8.95       && 394          && 58  		          && 40-80      \\ 
Cu\,Sn$_2$            && 8.90       && 140          && 25  		          && 150        \\ 
\hline\hline
\end{tabular} \end{center}
\end{table}

Aluminum or copper are favoured for electron facilities due to their high thermal conductivity. The synchrotron radiation can be absorbed directly by the vacuum chamber if the power line-density is not exceeding values of $\approx 100\,$W/m. With appropriate water cooling of the chamber a power load of several 10\,kW/m can be accepted. For intense synchrotron radiation fans with a shallow height the temperature profile should be simulated. The often quite elaborate beam pipe cross sections in combination with pumping and cooling ducts can be economically produced by continuous extrusion as shown in the example of Fig.\,\ref{fig:chamber}. More complex chambers are produced from solid blocks, which is associated with higher manufacturing cost. Neighboured chambers may be connected by aluminum flanges in combination with Helicoflex$^\RM{TM}$ gaskets. Alternatively aluminum/stainless steel transitions, for example made by explosion bonding, allow the usage of standard stainless steel Conflat$^\RM{TM}$ flanges. Overall the sealing of aluminum systems is not as reliable as the standard stainless steel system with copper gaskets. 

In some cases copper or copper alloy chambers are used to benefit from the even higher conductivity compared to Al. Examples include the accelerator facilities HERA-e, KEK-B and PEP. Brazing techniques are typically applied to join copper components and to connect stainless steel flanges. Massive water cooled copper blocks are often used as collimators or absorbers for synchrotron radiation at locations of high power density.   

\begin{figure} 
\centerline{\includegraphics[width=1.0\textwidth]{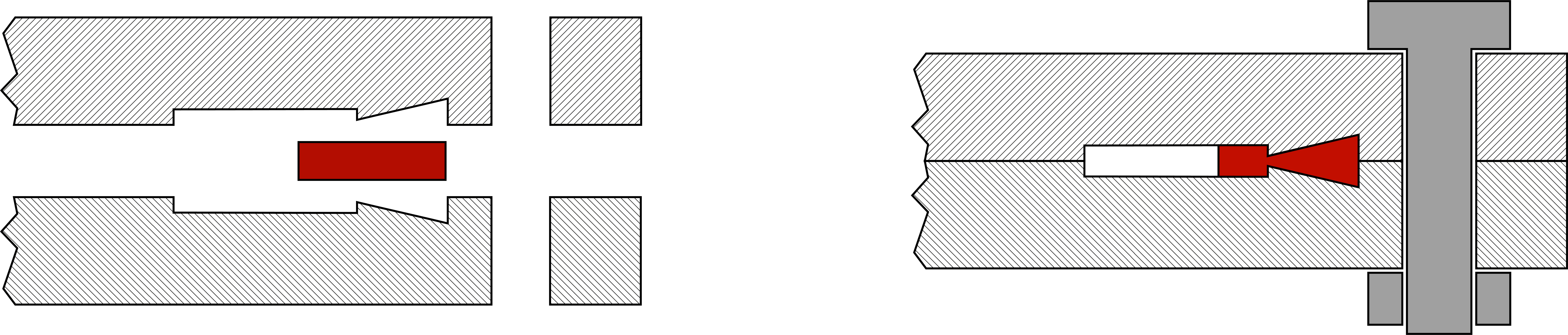}}
\caption{In a Conflat$^\RM{TM}$ metal sealed flange connection the copper seal is clamped between two stainless steel flanges that contain a knife edge.} 
\label{fig:conflat}
\end{figure}

For proton accelerators austenitic stainless steel has become the most widely used material. Beam pipes are often fabricated from seamless tubes with discrete pumps attached every few meters. For sealing usually welded stainless steel flanges and copper Conflat$^\RM{TM}$ gaskets are used. Only metal sealed flange connections allow to achieve leak rates compatible with UHV conditions and radiation resistance at the same time. A concept sketch of the Conflat system is shown in Fig.\,\ref{fig:conflat}.

For accelerators with superconducting magnets the vacuum chamber is often also operated at low temperature, since there is no room for thermal insulation in the tight space between the magnet coils. Examples include the proton ring of HERA/DESY, RHIC at Brookhaven and LHC at CERN. The~cool down over a large temperature range from room temperature to an operating temperature of, for example, $4.5\,$K presents another challenge for the beam vacuum system. A system of bellows and space for expansion must be foreseen to handle the mechanical contraction. Usually the beam pipes are made from a type of stainless steel with the inner surface coated by copper to enhance the thermal conductivity. 

For electrical feedthroughs into vacuum ceramics are used as insulators. In some cases ceramics are used for entire vacuum chambers. This is necessary if the chamber is to be placed in rapidly changing magnetic fields, and in a metallic chamber strong eddy currents would be induced. Applications include for example kicker magnets and rapid cycling synchrotrons \cite{csns}. Beam chambers inside particle physics detectors should influence particles generated at the interaction point as less as possible. For this purpose very thin wall tubes are used, made from low Z material, such as aluminum, beryllium or carbon fiber materials. For exit windows of beams similar solutions are common.

The beam chamber carries an image current that acts back on the beam and may deteriorate beam quality properties like emittance or energy spread. To minimize such effects the chamber should carry the~image current with low resistance, avoiding geometric discontinuities. Variations of the vacuum chamber cross section should be applied gradually using tapered sections. Bellows and openings for pumping ports must be electrically shielded. In bellows often a set of flexible, sliding spring contacts of copper-beryllium alloys is installed around their circumference. Perforated electric screens with circular holes of some millimeters diameter or longitudinal slits are used for pump ports. Of course such electrical shielding measures result in a reduction of the effective pumping speed. Also for gate valves in the open state, gaps have to be covered by spring contacts that move with the gate mechanism. Inappropriate shielding of transverse openings will not only affect the beam quality, but may  also result in the excitation of trapped RF modes, leading to a strong heating effect for vacuum components.

\begin{figure} 
\centering\includegraphics[width=0.85\textwidth]{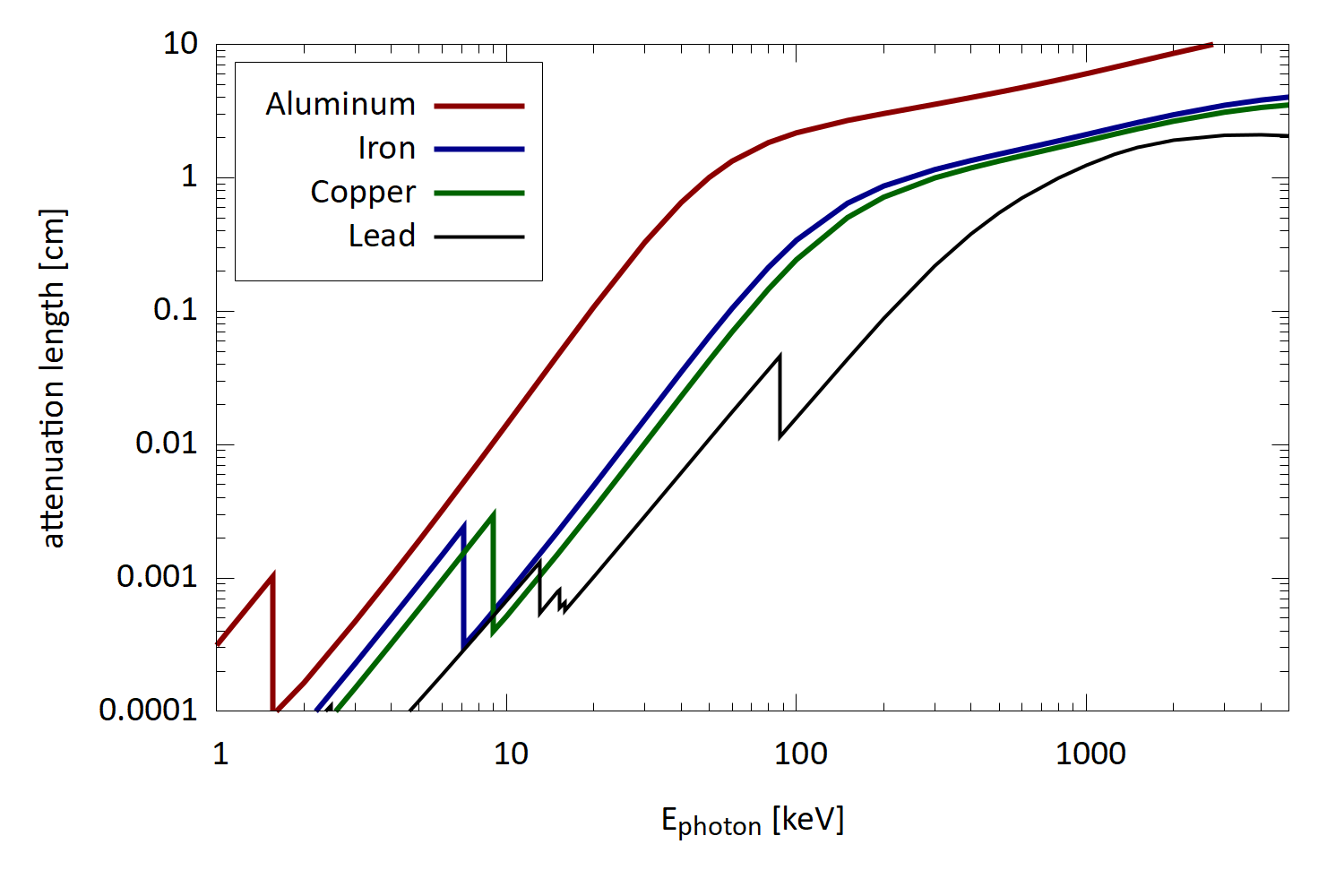}
\caption{Attenuation length for X-rays of commonly used materials for vacuum vessels and shielding as a function of photon energy.}
\label{fig:att}
\end{figure}

Accelerator components installed in the vicinity of the beam pipe should be radiation resistant and must withstand corrosive atmospheres produced by the primary radiation. This includes cables and electronics, which must either be properly chosen or shielded. For high energy electron/positron storage rings and synchrotron radiation facilities the generated X-rays may cause problems, and often the beam chamber is wrapped into a lead shield to absorb as much radiation as possible. Figure\,\ref{fig:att} shows the~attenuation of X-rays in various materials as function of the photon energy. Using beam energy and bending radius the critical photon energy may be estimated with the expression: 

\begin{equation}
E_c [\RM{keV}] \approx 2.218 ~ \frac{E^3 [\RM{GeV}^3]}{\rho [\RM{m}]}.
\label{crit}
\end{equation}

\section{Summary}
The vacuum system presents a challenging technical aspect of each particle accelerator. To achieve required beam lifetimes and beam quality, and to minimize unwanted losses of high energy particles associated with radio-activation and damage to components, vacuum systems must be carefully designed to provide the adequate low gas densities.

Steps for designing an accelerator vacuum system can be roughly categorized into three groups. The first step involves an evaluation of the {\bf gas sources} in an accelerator. This includes outgassing rates for surfaces and beam induced outgassing dynamics, such as synchrotron radiation, electron cloud, heavy ion bombardment to name a few. Certain accelerator designs require quick handling of activated components thus involving relatively leaky inflating seals. In such cases leak rates must be considered. In practice a clean and baked stainless steel surface might exhibit an outgassing rate of $q_0 = 10^{-11}\,$mbar\,l\,s$^{-1}$\,cm$^{-2}$. In the presence of synchrotron radiation and after a reasonable time of conditioning the outgassing is still dominated by releasing at least $\eta = 10^{-6}$ gas molecules per photon incident on the chamber wall.

The second step covers the definition of the {\bf target residual gas pressure and composition} (more precisely the gas density). Physics effects of beam gas interaction and the resulting performance degradation must be considered for this purpose. To give a couple of examples the acceptable dynamic gas pressure in an electron storage ring is in the order of $10^{-8}\,$mbar with a typical composition of $\sfrac{3}{4}$ H$_2$ and $\sfrac{1}{4}$ CO. A proton cyclotron as a single pass accelerator is more forgiving and a pressure of $10^{-6}$\,mbar is sufficient.

The third step involves then to lay out the vacuum system in terms of geometry, installed {\bf pumping speed} and perhaps more complex technical measures like surface coating to achieve the required vacuum quality under operating conditions. Types of pumps may include turbo pumps, ion sputter pumps, NEG coating or cryo pumps for large recipients. A typical electron storage ring may be equipped with $S = 100$\,l/s ion sputter pumps at a distance of 5\,m. Today software can be used to compute the pressure profile using Monte Carlo methods or numerical solutions of the diffusing Eq. (\ref{eq:diffeq}). 
In addition to such conceptual considerations the design of vacuum systems requires a lot of engineering. Keywords include: UHV compatible materials and materials preparation, mechanical stability, thermo-mechanical problems under heat load, pumps, gauges, flange systems and valves.

\end{document}